\documentclass[10pt, conference, compsocconf]{IEEEtran}
\IEEEoverridecommandlockouts
\usepackage{amsmath}
\usepackage{amssymb}
\usepackage{lipsum}
\usepackage[font=footnotesize, labelfont=bf]{caption}
\usepackage{listings}
\usepackage{gensymb}
\usepackage{fancyhdr}
\usepackage{graphicx}

\begin{document}
%
\title{Possibilities of Recursive GPU Mapping for Discrete Orthogonal Simplices}


\author{
\IEEEauthorblockN{Crist\'obal A. Navarro}
\IEEEauthorblockA{Institute of Informatics,\\
Universidad Austral de Chile\\
Valdivia, Chile\\
Email: cnavarro@inf.uach.cl}
\and
\IEEEauthorblockN{Benjam\'in Bustos}
\IEEEauthorblockA{Department of Computer Science (DCC)\\
University of Chile, Santiago, Chile\\
Email: bbustos@dcc.uchile.cl}
\and
\IEEEauthorblockN{Nancy Hitscheld}
\IEEEauthorblockA{Department of Computer Science (DCC)\\
University of Chile, Santiago, Chile\\
Email: nancy@dcc.uchile.cl}
}

\maketitle

\begin{abstract}
    The problem of parallel thread mapping is studied for the case of discrete
    orthogonal $m$-simplices. The possibility of a $O(1)$ time recursive
    block-space map $\lambda: \mathbb{Z}^m \mapsto \mathbb{Z}^m$ 
    is analyzed from the point of view of parallel space efficiency and
    potential performance improvement. The $2$-simplex and $3$-simplex are
    analyzed as special cases, where constant time maps are found, providing a
    potential improvement of up to $2\times$ and $6\times$ more efficient than
    a bounding-box approach, respectively.  For the general case it is shown
    that finding an efficient recursive parallel space for an $m$-simplex
    depends of the choice of two parameters, for which some insights are
    provided which can lead to a volume that matches the $m$-simplex for
    $n>n_0$, making parallel space approximately $m!$ times more efficient than a
    bounding-box. 
\end{abstract}


%
\IEEEpeerreviewmaketitle

\section{Introduction}
The field of GPU computing has become a well established research area in the
last ten years \cite{4490127, Nickolls:2010:GCE:1803935.1804055, navhitmat2014}
thanks to the high performance of programmable graphics hardware and the release
of a generic GPU programming model, being CUDA \cite{nvidia_cuda_guide} and
OpenCL \cite{opencl08} the most known implementations.  In the GPU programming
model there are three constructs\footnote{This work follows the naming scheme by
Nvidia CUDA. OpenCL chooses different names for these constructs; (1)
work-element, (2) work-group and (3) work-space, respectively.} that allow the
execution of highly parallel algorithms; (1) thread, (2) block and (3) grid.
Threads are the smallest elements and they are in charge of executing the
instructions of the GPU kernel.  A block is an intermediate structure that
contains a set of threads organized as an Euclidean box.  Blocks provide fast
shared memory access as well as local synchronization for all of its threads.
The grid is the largest construct of all three and it keeps all blocks together
spatially organized for the execution of a GPU kernel.  These three constructs
play an important role when mapping the execution resources to the problem
domain.

\begin{figure}[ht!]
\centering
\includegraphics[scale=0.13]{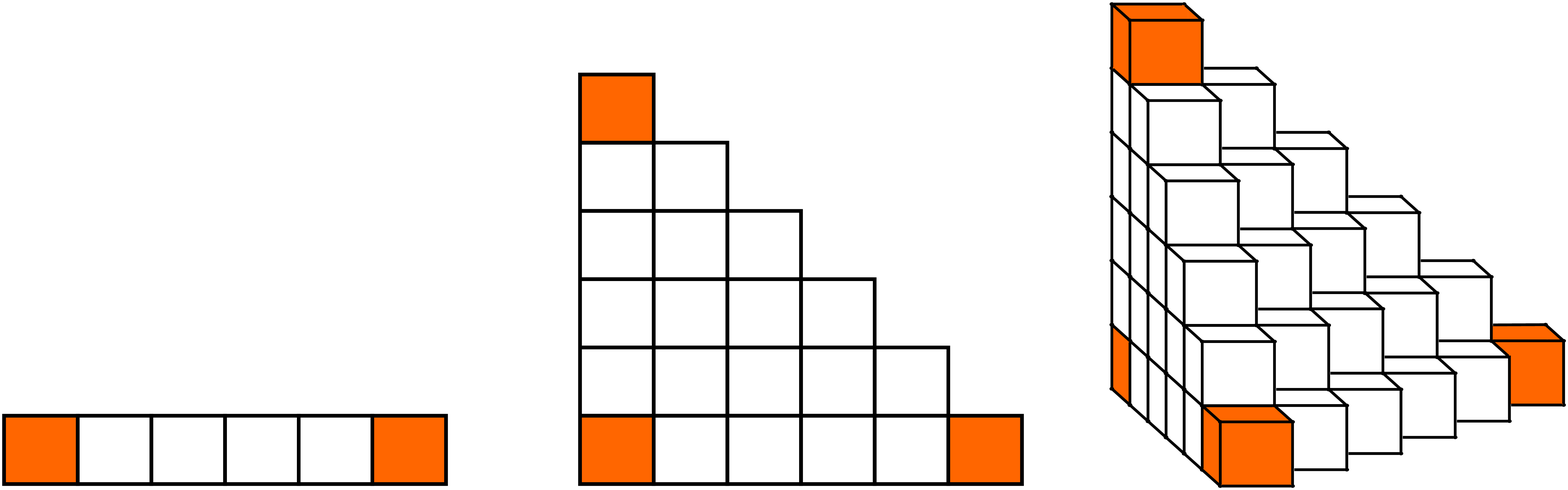}
\caption{Discrete orthogonal $m$-simplices up to $m=3$ dimensions.}
\label{fig_dom-simplex}
\end{figure}

For every GPU computation there is a stage where threads are mapped from
parallel to data space. A map, defined as $f: \mathbb{Z}^k \rightarrow
\mathbb{Z}^m$, transforms each $k$-dimensional point $x=(x_1, x_2, ..., x_k)$ in
parallel space into a unique $m$-dimensional point $f(x) = (y_1, y_2, \cdots,
y_m)$ in data space. GPU parallel spaces are defined as orthotopes
$\Pi_n^m$ in $m=1,2,3$ dimensions\footnote{Higher dimensional orthotopes can be still
be represented by linearizing to a three-dimensional one.}. A known
approach for mapping threads is to build a \textit{bounding-box} (BB) type
of orthotope, sufficiently large to cover the data space and map threads using
the identity $f(x) = x$. Such map is highly convenient and efficient for the
class of problems where data space is also defined by an orthotope; such as
vectors, tables, matrices and box-shaped volumes.  But there is a different
class of problems where data space follows a discrete orthogonal $m$-simplex
organization (see Figure \ref{fig_dom-simplex}).

\begin{figure}[ht!]
\centering
\includegraphics[scale=0.12]{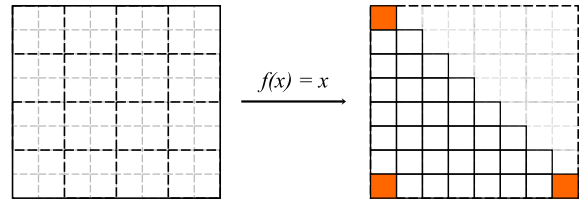}
\caption{For $m=2$, the bounding-box strategy generates a parallel space $P_2$ 
that approaches $\sim 2\times$ the required volume.}
\label{fig_bb_strategy}
\end{figure}

Problems such as the Euclidean distance matrix (EDM) \cite{5695222,
Li:2010:CME:1955604.1956601, Man:2011:GIC:2117688.2118809}, collision detection
\cite{AvrilGA12}, adjacency matrices \cite{kepner2011graph}, cellular automata
simulation on triangular/tetrahedral spatial domains \cite{ConwaysLife}, matrix
inversion \cite{Ries:2009:TMI:1654059.1654069}, \textit{LU/Cholesky}
decomposition \cite{springerlink_gustavson} and the \textit{n-body} problem
\cite{DBLP:journals/corr/abs-1108-5815, Bedorf:2012:SOG:2133856.2134140,
Ivanov:2007:NPT:1231091.1231100}, among others, follow the shape of a 
discrete orthogonal $2$-simplex, $\Delta_n^2$, with a volume of
$V(\Delta_n^2)=n(n+1)/2 \in \mathcal{O}(n^2)$. The default \textit{bounding-box}
(BB) approach turns out to be inefficient because the volume of its parallel
space, $V(\Pi_n^2)$, produces $n(n-1)/2 \in \mathcal{O}(n^2)$ unnecessary
threads (see Figure \ref{fig_bb_strategy}).

\begin{figure}[ht!]
\centering
\includegraphics[scale=0.14]{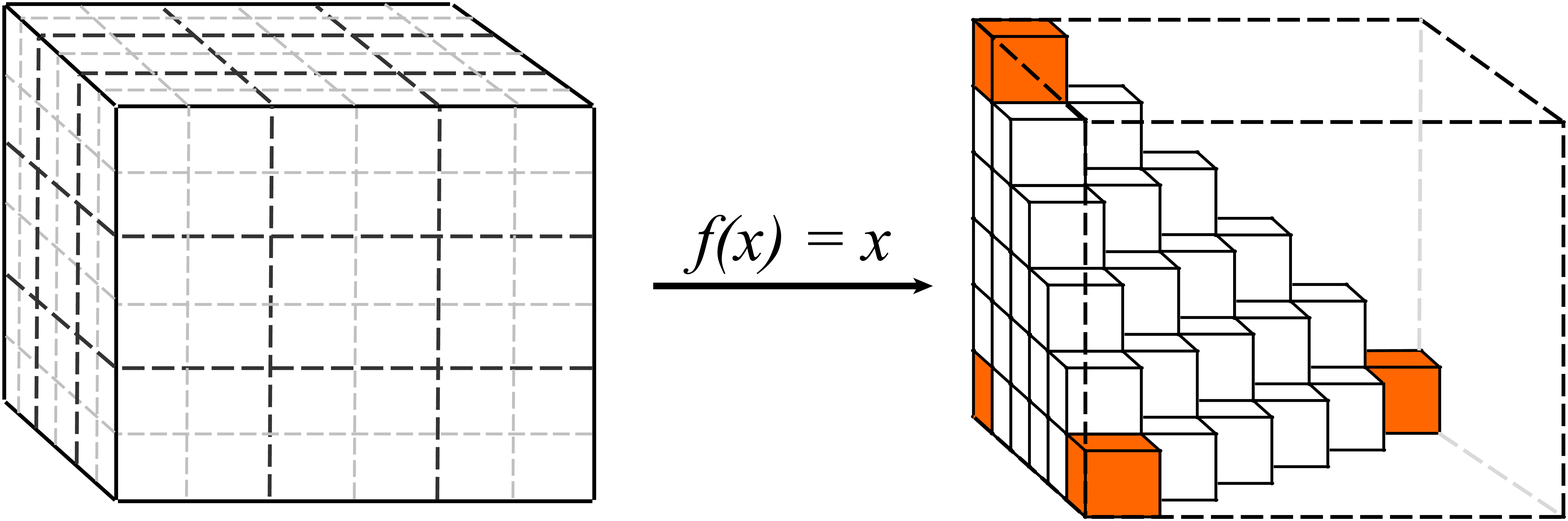}
\caption{Bounding-box approach mapped to a discrete orthogonal tetrahedron.}
\label{fig_bb_strategy_tetrahedron}
\end{figure}

Problems such as the triple-interaction n-body problem
\cite{Koanantakool:2014:CCP:2683593.2683634} and triple correlation analysis
\cite{1992Huckaby} are represented with a discrete orthogonal $3$-simplex.  In
the $3$-simplex class, data space has a size of $V(\Delta_n^3)=n(n+1)(n+2)/6 \in
\mathcal{O}(n^3)$ elements, organized in a tetrahedral way.  Once again, the
default \textit{bounding-box} (BB) approach is inefficient as it generates a
parallel volume $V(\Pi_n^3)$ with $\mathcal{O}(n^3)$ unnecessary threads (see
Figure
\ref{fig_bb_strategy_tetrahedron}).

In general, an orthogonal $m$-simplex is by definition an $m$-dimensional
polytope where its facets define a convex hull, with one vertex having all of
its adjacent facets orthogonal one to each other. A discrete orthogonal
$m$-simplex, denoted as $\Delta_n^m$, is the analog of the continuous one, but
volumetric and composed of a finite number of discrete elements $\vec{x} =
\{x_1, x_2, ..., x_m\}$ that can be characterized as
\begin{equation}
    \Delta_n^m \equiv \{\vec{x} \in \mathbb{Z}_+^m | 0 \le x_i \le n \land x_1 + x_2 + ...
x_m \le n\}.
\end{equation}
which establishes an upper bound for the absolute Manhattan distance from any
element $\vec{x}$ to the orthogonal corner of the $m$-simplex. The expression for the
volume of an $m$-simplex is well defined by the Simplicial polytopic numbers
\begin{equation}
    V(\Delta_n^m) = {{n+m-1}\choose{m}} = \frac{n(n+1)...(n+m-1)}{m!}
    \label{eq_volume}
\end{equation}
which can be proved using an induction \cite{1054599} on the fact that the
volume of $\Delta_n^{m+1}$ is the sum of the volumes of $n$ stacked
$m$-simplexes of lengths $\{1, 2, 3, ..., n\}$, \textit{i.e.},
\begin{equation}
    V(\Delta_n^{m+1}) = \sum_{i=1}^n{V(\Delta_i^m)}
    \label{eq_volume_enum}
\end{equation}
which when combined with the properties of sums of binomial coefficients, leads
to expression (\ref{eq_volume}).
When using a bounding-box approach, the fraction of extra volume of $V(\Pi_n^m)$
that lies outside of the $m$-simplex approaches to
\begin{equation}
    \lim_{n\to\infty} \alpha(\Pi, \Delta)_n^m = \frac{V(\Pi_n^m)}{V(\Delta_n^m)}
    - 1 =  m!-1
\end{equation}
making it an inefficient approach for large $n$ as $m$ increases.


A natural enumeration approach can be used by expanding expression
(\ref{eq_volume_enum}) and indexing the elements in a linear way. Such approach 
allows to formulate a map of the form $g:\mathbb{Z}^1 \mapsto \mathbb{Z}^m$
with $V(\Pi_n^m) = V(\Delta_n^m)$.
Although $g(\vec{x})$ may be computable by arithmetic and elementary functions, its
complexity increases with $m$ as it requires the solution of an $m$-th order
equation. Furthermore, the method is limited to $m \le 4$ as no analytical
solutions exist for polynomials of $m \ge 5$. It is of interest then to find a
different kind of map, free of such problems.

The limitations of the enumeration principle can be overcome, in great part, by
taking advantage of the dimensionality available in the parallel space. Although
parallel spaces in GPU cannot have a geometry different from an orthotope, they
can be $m$-dimensional which makes them topologically equivalent to an
$m$-simplex.  Finding an homeomorphism of the form $\lambda: \mathbb{Z}^m
\mapsto \mathbb{Z}^m$ would produce zero dimensional distance between parallel
and data spaces which would free it from the computation of $m$-th roots. 

This work presents a study of the possibilities of recursive GPU mapping of
thread-blocks onto $m$-simplices. A dedicated analysis is devoted to the special
cases of $2$-simplex and $3$-simplex, where $O(1)$ time maps are found and
described, offering a space improvement of $2\times$ and $6\times$,
respectively, that results in a potential performance improvement given that no
$m$-roots are required.  For general $m$ it is shown that building an efficient
set of recursive orthotopes requires finding optimal values for the reduction
factor $r$ and the arity $b$. Values for both parameters are analyzed, giving
the possibility to build highly tight recursive volumes for $n \ge n_0$, making
an improvement of $m!$ in parallel space efficiency with respect to the
bounding-box approach.   

The rest of the manuscript presents related work (Section
\ref{sec_related-work}), a formal definition and analysis of
$\lambda(\vec{\omega})$ (Section \ref{sec_formulation}) for the different cases
and finally the main results are discussed including future work (Section
\ref{sec_discussion}).

\section{Related Work}
\label{sec_related-work}
Ying \textit{et. al.} have proposed a GPU implementation for parallel
computation of DNA sequence distances \cite{Ying:2011:GDD:2065356.2065583} which
is based on the Euclidean distance maps (EDM), a problem in the $2$-simplex
class. The authors mention that the problem domain is indeed symmetric and they
do realize that only the upper or lower triangular part of the interaction
matrix is sufficient. Li \textit{et. al.} \cite{Li:2010:CME:1955604.1956601}
have also worked on GPU-based EDMs on large data and have also identified the
symmetry involved in the computation.
 
Jung \textit{et. al.} \cite{Jung2008} proposed packed data structures
for representing triangular and symmetric matrices with applications to LU and
Cholesky decomposition \cite{springerlink_gustavson}. The strategy is based on
building a \textit{rectangular box strategy} (RB) for accessing and storing a
triangular matrix (upper or lower). Data structures become practically half the
size with respect to classical methods based on the full matrix. The strategy
was originally intended to modify the data space (\textit{i.e.,} the matrix),
however one can apply the same concept to the parallel space.

Ries \textit{et. al.} contributed with a parallel GPU method for the triangular
matrix inversion \cite{Ries:2009:TMI:1654059.1654069}.  The authors identified
that the parallel space indeed can be improved by using a \textit{recursive
partition} (REC) of the grid, based on a \textit{divide and conquer} strategy.
The approach takes $O(\log_2(n))$ time by doing a balanced partition of the
structure, from the orthogonal point to the diagonal.

Q. Avril \textit{et. al.} proposed a GPU mapping function for collision
detection based on the properties of the \textit{upper-triangular map}
\cite{AvrilGA12}. The map is a thread-space function $u(x) \rightarrow (a, b)$,
where $x$ is the linear index of a thread $t_x$ and the pair $(a,b)$ is a unique
two-dimensional coordinate in the upper triangular matrix. Since the map works
in thread space, the map is accurate only in the range $n \in [0, 3000]$ of 
linear problem size.

Navarro, Hitschfeld and Bustos have proposed a block-space map function for
$2$-simplices and $3$-simplices \cite{DBLP:conf/hpcc/NavarroH14,
CLEI-2016-navarro}, based on the solution of an $m$ order equation that is
formulated from the linear enumeration of the discrete elements. The authors
report performance improvement for $2$-simplices. For the $3$-simplex case, the
mapping technique is extended to discrete orthogonal tetrahedron, where the
parallel space usage can be $6\times$ more efficient. However the authors 
clarify that it is difficult to translate such space improvement into
performance improvement, as the map requires the computation of
several square and cubic roots that introduce a significant amount of overhead
to the process. From the point of view of data-reorganization, a succinct blocked
approach can be combined along with the block-space thread map, producing
additional performance benefits with a sacrifice of $o(n^3)$ extra memory. 

The present work proposes a new type of map $\lambda(\vec{\omega})$ that uses a
recursive organization of blocks but does not require multiple passes to map
threads onto the data space. Instead, it maps all blocks directly to the data space
by using a flat expression of lower computational cost than the
non-linear maps proposed in the past, which were based on the enumeration principle
\cite{AvrilGA12, DBLP:conf/hpcc/NavarroH14, CLEI-2016-navarro}.

\section{Formulation of $\lambda(\vec{\omega})$}
\label{sec_formulation}
\textit{Note: for
practical purposes, a discrete orthogonal $m$-simplex will be just referred as an
$m$-simplex.}

The formulation of $\lambda(\vec{\omega})$ begins by considering the special cases
$m=2,3$, where the mapping is graphically represented.  Case $m=1$ is not
considered as both orthotopes and simplices match in geometry. 

\subsection{Mapping to $2$-Simplices}
\begin{figure}[ht!]
\centering
\includegraphics[scale=0.105]{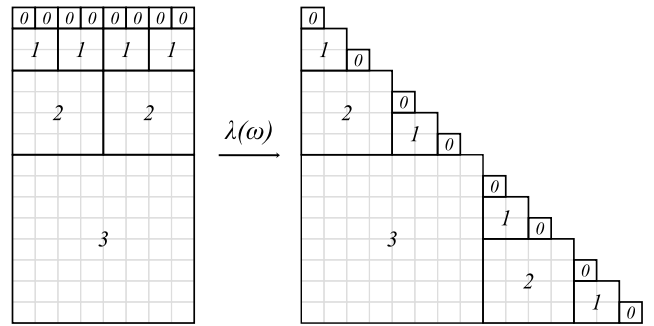}
\caption{Both $\Pi_n^2$ and $\Delta_n^2$ can be defined by a recursive set
    $S_{n+1}^2$.}
\label{fig_m2-ortho-simplex-map}
\end{figure}

For a $2$-simplex the volume of $\Delta_n^2$ is given by the triangular numbers
\begin{equation}
    V(\Delta_n^2) = \frac{n(n+1)}{2}
\end{equation}
An orthotope $\Pi_n^2$ can be subdivided by a set $S_n^2$ of
self-similar sub-orthotopes with a recursive structure, giving a volume of 
\begin{equation} 
    V(\Pi_n^2) = V(S_n^2) = \Big(\frac{n}{2}\Big)^2 + 2V(S_{n/2}^2)
\end{equation}
with a boundary condition of $V(S_2^2) = 1$ and $n = 2^k$ with $k \in
\mathbb{Z}_+$. Its expanded form produces the sum
\begin{alignat}{2}
    V(S_n^2) &= 2^0\Big(\frac{n}{2^1}\Big)^2 + 2^1\Big(\frac{n}{2^2}\Big)^2
    +\dots +2^{\log_2 n}\Big(\frac{n}{2^{\log_2 n}}\Big)^2\\
    &= \frac{n^2}{2} \sum_{i=1}^{\log_2 n} \Big(\frac{1}{2^{i}}\Big).
\end{alignat}
where its reduction via the geometric series $\sum_{i=0}^{k}{a^i} =
\frac{a^{k+1} - 1}{a-1}$, results in
\begin{alignat}{2}
    V(S_n^2) &= \frac{n^2}{2} \Bigg(-1 +  \sum_{i=0}^{\log_2 n}
    {\Big(\frac{1}{2}\Big)^i}\Bigg)\\
    &= \frac{n^2}{2} \Bigg( -1 + \frac{ (1/2)^{\log_2{n} + 1} - 1 }{1/2 - 1}
    \Bigg) \\
    \label{eq_m2}
    &= \frac{n(n-1)}{2} = \Delta_{n-1}^m.
\end{alignat}
The result from expression (\ref{eq_m2}) is equivalent to 
\begin{equation}
    V(S_{n}^2) + n = V(S_{n+1}^2) =  V(\Delta_n^2).
\end{equation}
which means that set $S_{n+1}^2$ can be organized both as an orthotope $\Pi_n^2$
as well as a $2$-simplex $\Delta_n^2$. Therefore, a proper block-space
homeomorphism $\lambda: \mathbb{Z}_+^2 \mapsto \mathbb{Z}_+^2$ could map
$\Pi_n^2$ onto $\Delta_n^2$ and provide an improvement in parallel space
efficiency, as shown in Figure \ref{fig_m2-ortho-simplex-map}.
 
    Let $\vec{\omega} = (w_x,w_y)$ be a block of threads in parallel space $\Pi_n^2$ (each
    block is illustrated as a gray lined square in Figure
    \ref{fig_m2-ortho-simplex-map}) located at $(x,y)$, with the origin at the
    top-left corner. The value $w_y$ of a block can be used to obtain the
    recursion level, \textit{i.e.}, $\lfloor \log_2 y \rfloor$, which is used to
    define the starting height value $b = 2^{\lfloor \log_2 y \rfloor}$ for the
    type of orthotope whom $w_{x,y}$ belongs to.  The value $q = \lfloor (w_x/b)
    \rfloor$ provides a way to know which of the sub-orthotopes of the same level
    $w_{x,y}$ belongs to.  The combination of these parameters allows the
    formulation of the homeomorphism
\begin{equation}
    \label{eq_map2d}
    \lambda(\vec{\omega}) = (w_x + qb, w_y + 2qb)
\end{equation}
    that maps in $O(1)$ time, which is a considerable improvement over the
    recursive triangular map that requires $O(log_2(n))$ recursive steps
    \cite{Ries:2009:TMI:1654059.1654069}, even when it is still based on a
    recursive organization of elements. Additionally, since blocks have a constant size of
    $\rho^2 \ll n$, with $\rho$ the number of threads in each
    dimension\footnote{For simplicity, equal block dimensions have been chosen,
    although the results are not limited to this assumption.}, the extra number
    of threads is no greater than $n\rho^2 \in o(n^2)$.

    The computation of $\lambda(\vec{\omega})$ requires a small number of arithmetic 
    operations and only two elementary functions. The function $\lfloor log_2(y)
    \rfloor$ can be computed using the binary relation
    \begin{equation}
        \lfloor log_2(y) \rfloor = b - clz(y)
    \end{equation}
    where $b$ is the number of bits of the word and $clz(x)$ counts the number
    of leading zero bits of $y$. The exponential $2^{\lfloor \log_2(y) \rfloor}$
    can be computed using
    \begin{equation}
        2^{\lfloor \log_2(y) \rfloor} = 2 << (b - clz(y))
    \end{equation}
    Considering that the two elementary functions can be computed using
    bit-level operations, and that the parameters are re-used by registers, it
    is expected that the parallel space improvement from $O(n^2)$ to $O(n)$
    unnecessary threads (\textit{i.e.}, the number of unnecessary threads over
    the diagonal is no greater than $\rho^2n \in O(n)$) can indeed result in a
    significant performance improvement, which for the case of triangles is in
    the range of $0 \le I \le 2$ \cite{DBLP:conf/hpcc/NavarroH14}.  Moreover,
    since no square roots are required, $\lambda(\vec{\omega})$ has the potential 
    to be faster than previous mapping techniques based on the analytic solution 
    of a quadratic equation \cite{AvrilGA12, DBLP:conf/hpcc/NavarroH14}.

    The analysis of $\lambda(vec{\omega})$ has assumed problems with sizes of
    the form $n = 2^k$. For any value of $n$, one can use any of the following approaches:
    \begin{enumerate}
        \item \textit{Approach $n$ from above}: build a single orthotope $\Pi_{n'}^2$, where $n' =
                2^{\lceil \log_2(n) \rceil}$ and filter out the threads outside
                the domain. This approach keeps simplicity at
                the cost of adding extra threads.
        \item \textit{Approach $n$ from below}: apply a set of orthotopes $\Pi_{n_1}^2, \Pi_{n_2}^2,
            \Pi_{n_i}^2, ...$, where $n_i = log_2\Big(n - \sum_{k=1}^{i-1}
            n_k\Big)$ for $i \ge 2$, $n_1 = \lfloor log_2(n) \rfloor$, plus a
            set of more simpler mappings for the sub-orthotopes that remain
            un-mapped at each level. This other approach does not add extra
            threads but adds complexity.
    \end{enumerate}
    
    Choosing one or the other can depend on the particular type of problem.
    Nevertheless, it is important to mention that in many cases, such as
    in physical simulations, it is possible to adapt the problem size to $n =
    2^k$, making it possible to use $\lambda(\vec{\omega})$ in its intended form.

\subsection{Mapping to $3$-Simplices}
\begin{figure}[ht!]
\centering
\includegraphics[scale=0.065]{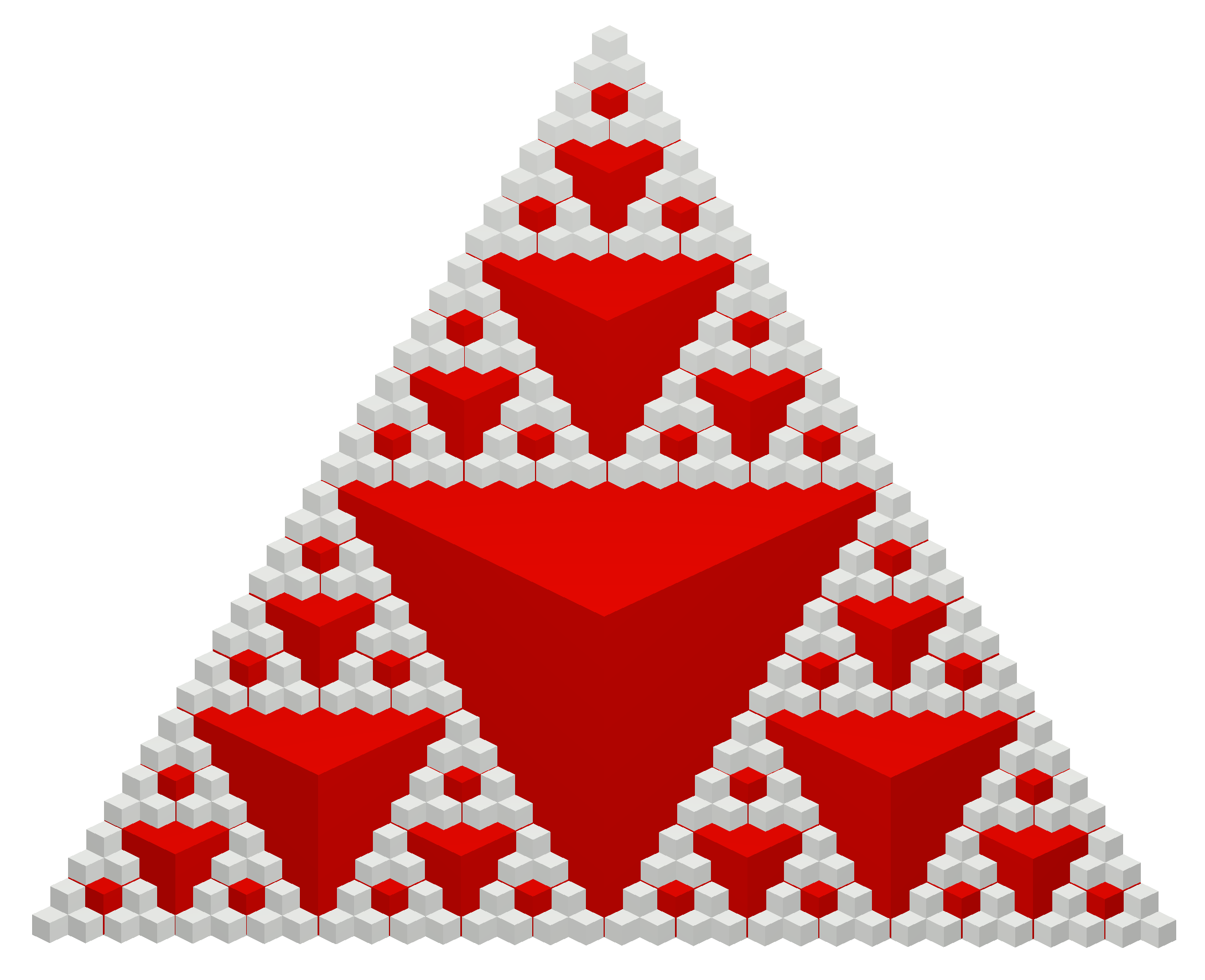}
\includegraphics[scale=0.065]{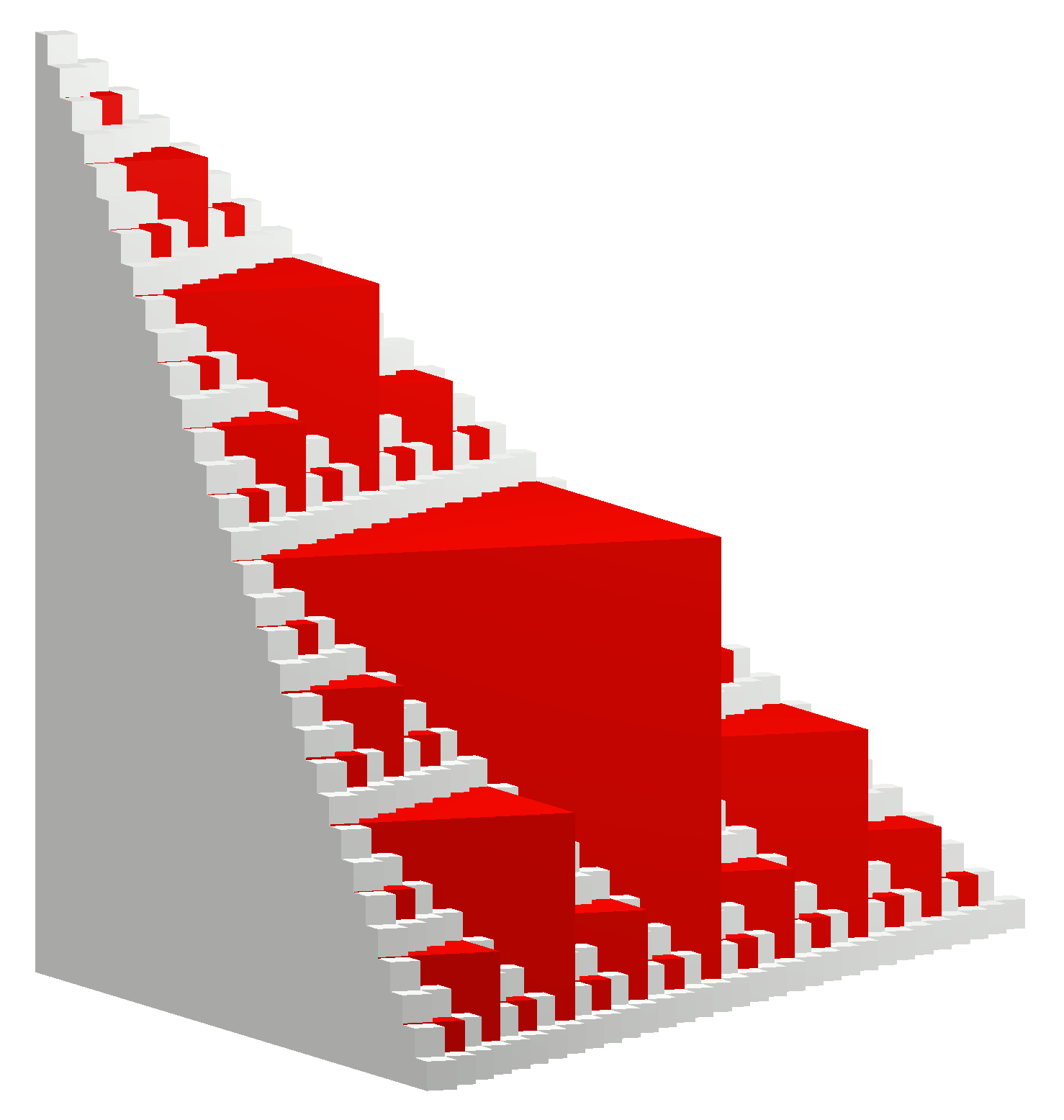}
\caption{Two different views of how the orthotope set (red) maps to the
tetrahedron (white cells) using an arbitrary number of recursions.}
\label{fig_m3-tetrahedron-binary-map}
\end{figure}

For a $3$-simplex of size $n$ per dimension, denoted as $\Delta_n^3$, its volume
is given by the tetrahedral numbers
\begin{equation}
    V(\Delta_n^3) = \frac{n(n+1)(n+2)}{6}.
\end{equation}
It is important to identify that there are multiple ways of extending the 
two-dimensional approach to three dimensions. 
One way to formulate $S_n^{m=3}$ is to extend the binary approach used 
in $2$-simplices, now to half-cubes with an arity of $\beta = 3$ for the
recursion, as the illustration of Figure \ref{fig_m3-tetrahedron-binary-map}. 

From the illustration, the red sub-volumes that form a structure similar to 
the Sierpinski gasket correspond to the parallel space that lies outside of the
tetrahedron.  It is relevant to know what is the volume of this fractal
structure, relative to the tetrahedron volume.  

The recursive orthotope set has the volume expression 
\begin{equation}
    V(S_n^3) = \Big(\frac{n}{2}\Big)^3 + 3V(S_{n/2}^3) = \frac{n^3}{3}
    \sum_{i=1}^{\log_2(n)} \Big(\frac{3}{2^3}\Big)^i
\end{equation}
where its reduction via geometric series is
\begin{alignat}{2}
    \label{eq_m3}
    V(S_n^3) &= \frac{n^3}{5} - 3^{\log_2(n)}.
\end{alignat}
In the infinite limit of $n$, the extra volume approaches to
\begin{equation}
    \lim_{n\to\infty} \alpha(S,\Delta)_n^m = \lim_{n\to\infty} \frac{ \frac{n^3}{5} -
3\log_2(n)}{ \frac{n(n+1)(n+2)}{6}} - 1 = \frac{1}{5}
\end{equation}
Considering that the extra volume constitutes no more than $20\%$ of the volume
of the tetrahedron, one can consider that this recursive strategy does not
suffer from significant extra volume problems in $m=3$. However, 
organizing the set $S_n^3$ into a single major orthotope $\Pi_n^3$ of dimensions
$(n-1) \times n/2 \times (n+1)/3$ to match $\Delta_n^3$ is not trivial as the
largest sub-orthotope is already greater than $(n+1)/3$ and each recursion adds
three sub-structures, leaving a gap when trying to close $\Pi_n^3$. Forcing the
sub-orthotopes to fit through deformation is neither an efficient approach, as
it would introduce greater complexity to the map $\lambda(\vec{\omega})$. For this 
reason mapping in $O(\log_2(n))$ recursive levels is re-considered for $3$-simplices,
as it is a practical approach that allows to keep the arithmetic computations
simple. The map $\lambda^{m=3}$ can be formulated as
\begin{alignat*}{3}
    \lambda(\vec{\omega}, \vec{c}, n)^{m=3} =  \varphi(\vec{\omega}, \vec{c})_{n/2}+
    \lambda(\vec{\omega}, \vec{c} + (\frac{n}{2}, 0, 0))_{n/2} + \\
    \lambda(\vec{\omega}, \vec{c} + (0, \frac{n}{2}, 0), \frac{n}{2}) +\\
    \lambda(\vec{\omega}, \vec{c} + (0, 0, \frac{n}{2}))_{n/2} +
\end{alignat*}
where $\vec{c}$ is the relative center and $\varphi(\vec{\omega}, \vec{c})_n = w +
\vec{c}$ for a cube of $n^3$ blocks. The map begins mapping the major cube of
$(n/2)^3$ blocks to the initial origin $\vec{c} = (0,0,0)$, then it recursively
calls three more maps with the corresponding new relative origins which are
located at the top and the sides of the cube. This process is repeated until the
smallest sized block is reached. In the end, the total number of map calls must
be at least
\begin{alignat}{2}
    \sum_{i=1}^{\log_2(n)} 3^i &\ge \frac{2^{\log_2(n) + 1} - 1}{3 - 1}  =
    \frac{n - 1}{2} = O(n).
\end{alignat}
Although the number of map calls is at least linear, in practice it turns out to be an
excessive number of parallel calls for the GPU computing model, which at the
present time can handle up to 32 concurrent kernels.  A more efficient map free 
of $O(n)$ recursive calls can be formulated by doing a small modification to the
strategy.

\subsection{Alternative Map for $3$-Simplices}
Although the previous map works in $O(\log_2(n))$ time, its main disadvantage is
the number of recursive calls to the map, making it unlikely to work efficient
when implemented on a GPU.

\begin{figure}[ht!]
\centering
\includegraphics[scale=0.06]{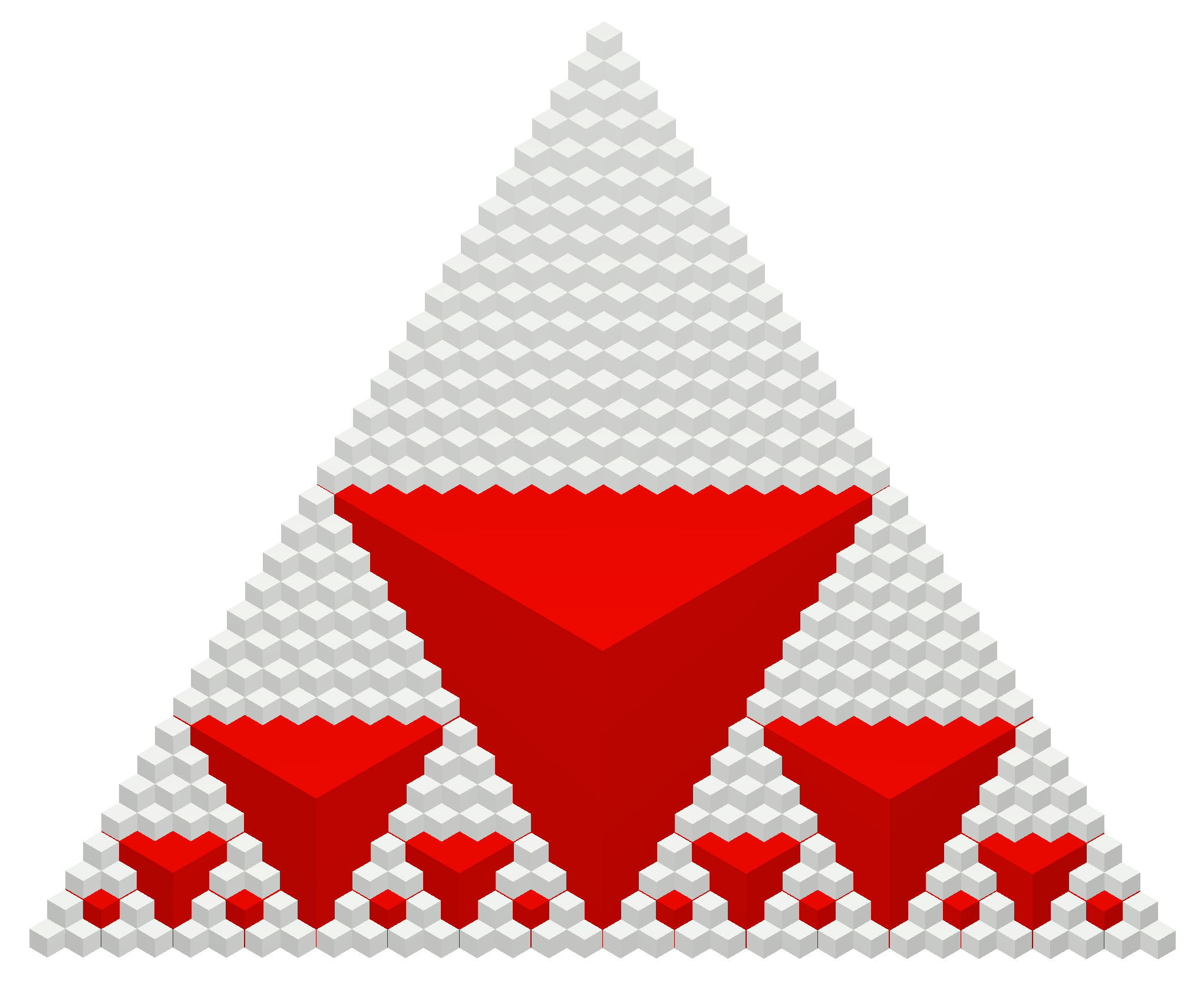}
\includegraphics[scale=0.06]{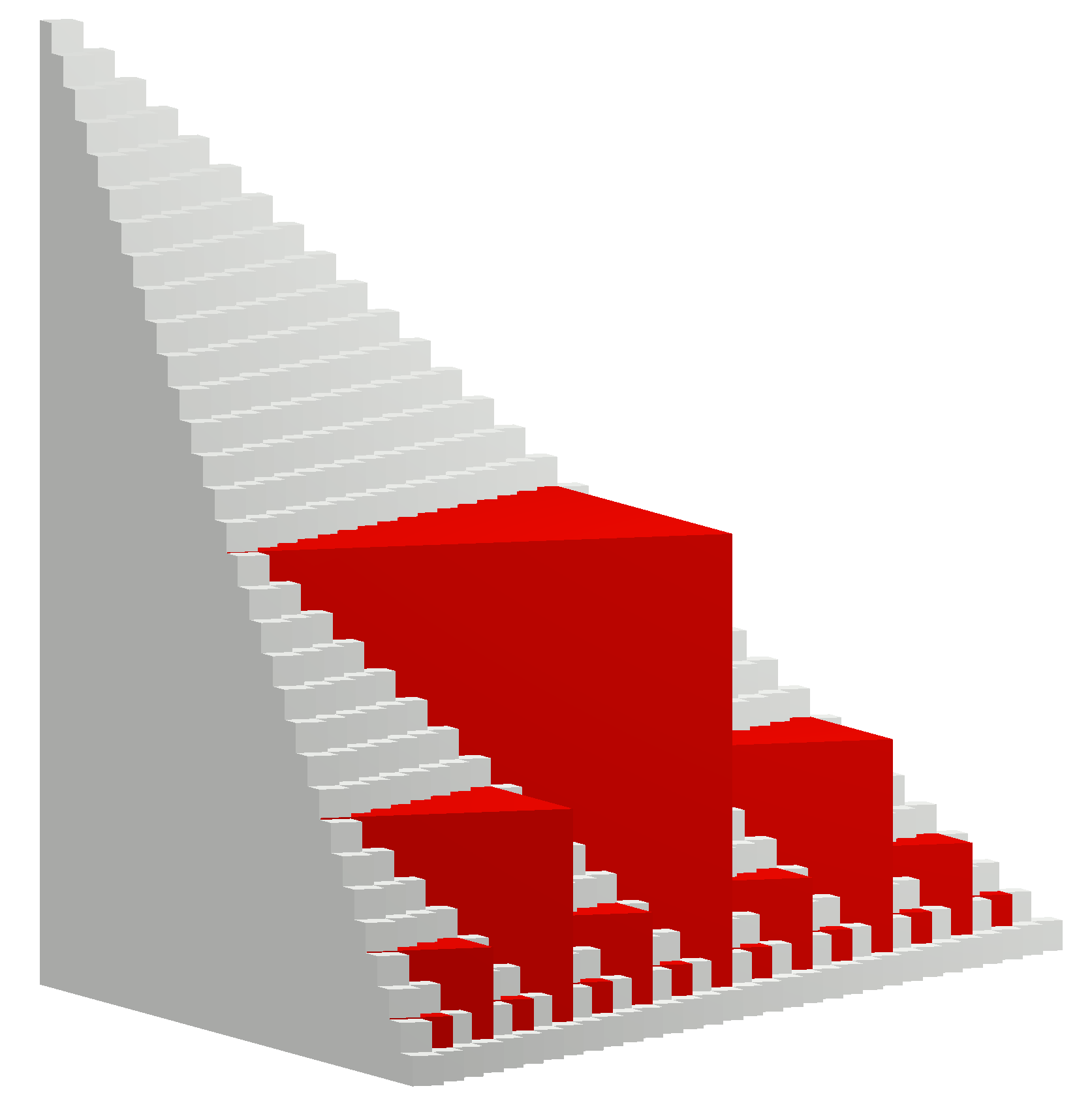}
\caption{Two different views of how the orthotope set (red) maps to the
tetrahedron (white cells) with only two recursion branches.}
\label{fig_optimization-m3}
\end{figure}

It is possible to modify the strategy and improve the efficiency of the parallel
space, as well as the map, by realizing that the recursive set $S_n^{3}$
can actually match the volume of the tetrahedron $\Delta_n^3$ by taking out one
of the recursion branches initially established. By doing this, the red
sub-tetrahedrons lying in the empty spaces of the Sierpinski gasket volume can
correspond to a unique uncovered sub-tetrahedron of data-space lying inside
$\Delta_n^3$. The process is done recursively, making it an
effective optimization.  The modified strategy is illustrated in Figure
\ref{fig_optimization-m3}.

\begin{figure}[ht!]
\centering
\includegraphics[scale=0.1]{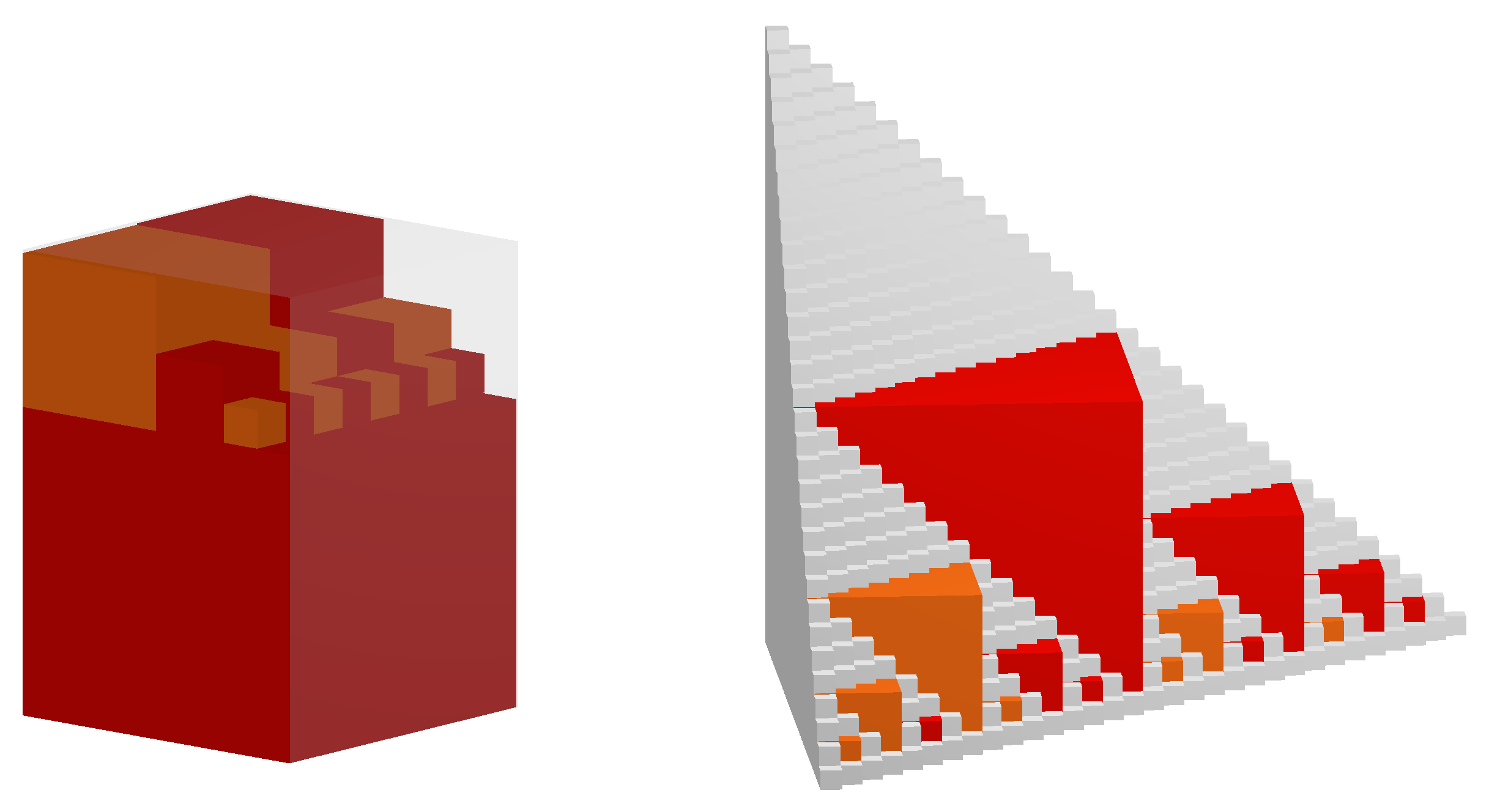}
\caption{Two different views of how the orthotope set (red) maps to the
tetrahedron (white cells) with only two recursion branches.}
\label{fig_map-fast-m3}
\end{figure}

With the new approach, the volume of the redefined set $S_n^3$ becomes
\begin{equation}
    V(S_n^3) = \Big(\frac{n}{2}\Big)^3 + 2V(S_{n/2}^3) = \frac{n^3}{2}
    \sum_{i=1}^{\log_2(n)} \Big(\frac{1}{4}\Big)^i
\end{equation}
which can be reduced using the geometric series
\begin{alignat}{2}
    \label{eq_m3}
    V(S_n^3) &= \frac{n^3 - n}{6} = V(\Delta_{n-1}^{3}).
\end{alignat}
As in the $2$-simplex map, the diagonal plane is not
considered, therefore the relation of data coverage is $V(S_{n+1}^3) =
V(\Delta_n^3)$. With this new organization it is now possible to build a $O(1)$ time map
free of the problems found in the original formulation and free of square roots.
The mapping works with a main orthotope of dimensions $V(\Pi_n^3) =
\Big(\frac{n}{2}\Big) \times \Big(\frac{n}{2}\Big) \times \frac{3(n-1)}{4}$ for
$x,y,z$, respectively. Figure \ref{fig_map-fast-m3} 
illustrates the new map.

Function $\lambda(\vec{\omega})$ assumes the origin of $\Pi_n^3$ at the bottom-right
corner from Figure \ref{fig_map-fast-m3} and the the origin of $\Delta_n^3$ at
the bottom right corner too, with the axes aligned to its orthogonal sides.  The
mapping begins by moving the main sub-orthotope of $(n/2)^3$ directly onto the
center of the tetrahedron with a simple map of the type
\begin{equation}
    h(\vec{\omega}) = \vec{\omega} + (0,\frac{n}{2},0).
\end{equation}
At the same time the rest of the sub-orthotopes map as
\[
\lambda(\vec{\omega}) = 
    \begin{cases} 
        (\omega_x + qb, \omega_y + 2qb, \omega_z - n/2),\ inside\\
        (b(1+2q) - \omega_x, 2b(1+q) - \omega_y,2b-\omega_z + \frac{n}{2}),\\
        diagonal \lor outside
    \end{cases}
\]

Parameters $q, b$ have the same definitions as in the $2$-simplex map and the total cost 
is constant in time, \textit{i,e,}, $T(h(\omega)) +
T(\lambda(\vec{\omega})) = O(1)$ even if done in sequence. 
The extra volume introduced by this approach is 
\begin{equation}
    \alpha(\Pi,\Delta)_n^3 = \frac{V(\Pi_n^{m=3})}{V(\Delta_n^{m=3})} - 1 =
    \frac{\frac{3n^2(n-1)}{16}}{\frac{(n-1)n(n+1)}{6}} - 1= \frac{2}{16}
\end{equation}
making $\Pi_n^{m=3}$ only $12.5\%$ larger than $\Delta_n^{3}$. Such amount of
extra volume constitutes a small fraction of the bounding-box that surrounds the
tetrahedron, which is practically $600\%$ the volume of $\Delta_n^m$ for large
$n$. For this reason, there is a potential performance improvement that can be
exploited by GPUs when using the optimized version of $\lambda(\vec{\omega})$ on $3$-simplices.

\subsection{Considerations for $m$-Simplices.}
The maps proposed for the $2$-simplex and $3$-simplex followed specific designs
for their corresponding dimensions. Although the maps take constant time for
both cases, it is important to note that for the $3$-simplex it was necessary
to introduce $12\%$ of extra parallel volume in order to fit the set $S_n^m$ on both
$\Pi_n$ and $\Delta_n$ and produce a single-pass map. When generalizing the
approach to $m$-simplices, it is important to first verify if $V(S_n^m) \ge
V(\Delta_n^m)$ satisfies as well as to find out how much extra space is
introduced.

The task is to build a set of recursively organized orthotopes $S_n^m$, where
the following satisfies $\forall \vec{x} \in \Delta_n^m, \vec{x} \in S_n^m$.
For general $m$, the volume of a set of recursive orthotopes $S_n^m$ is 
defined as
\begin{alignat}{2}
    V(S_n^m) &= (rn)^m + \beta V(S_{rn}^m) = (rn)^m \sum_{i=0}^{
        \log_{1/r}(n) - 1} (\beta r^m)^i
\end{alignat}
where $r$ is the scaling factor and $\beta$ the arity of the recursion. Applying
the geometric series, the expression becomes
\begin{alignat}{2}
    V(S_n^m) &= (nr)^m \Bigg( \frac{(\beta r^m)^{\log_{1/r}(n)} - 1}{\beta r^m - 1}\Bigg)\\
    \label{eq_generic-m}
            &= \frac{n^m - \beta^{\log_{1/r}(n)}}{1/r^m -\beta}.
\end{alignat}
where at least $V(S_n^m) \ge V(\Delta_{n-1}^m)$  must hold. For $m=2$, it is possible
verify that setting $r=1/2$ and $\beta = 2$ leads to equations (\ref{eq_m2}) 
and (\ref{eq_m3}) for $m=2,3$, respectively. For $m=4$ the 
total volume is
\begin{equation}
    V(S_n^m) = \frac{n^4 - n}{14} > \frac{(n-1)n(n+1)(n+2)}{24},\ n \ge 2
\end{equation}
and for large $n$ the extra volume introduced approaches to $5/7$ of
$\Delta_n^m$. For large $n$ in higher dimensions, the recursive strategy of
using $r=1/2$, $\beta=2$ produces a fraction of extra volume of
\begin{equation}
    \lim_{n\to\infty}\alpha(S,\Delta)_n^m = \lim_{n\to\infty} \frac{\frac{n^m- n}{2^m - 2}}{ {{(n-1) + m -
    1}\choose{m}}} - 1= \frac{m!}{2^m - 2} - 1
\end{equation}
which makes it inefficient for higher dimensions, \textit{i.e.}, for $m=5$ and $m=7$ it
produces $3\times$ and $39\times$ the volume of $\Delta_n^m$. 

A more efficient set $S_n^m$ can be found by searching the right values for $r$ and
$\beta$ in order to satisfy $1/(r^m) - \beta = m!$ or at least approach it
from below. The restriction however is that the term $\beta^{\log_{1/r}(n)}$
needs to be positive and should not grow too fast as it has an impact on the
efficiency of the parallel space. 

For example, a value of $r=1/(m^{-1/m})$ produces the required $m!$,
making $\beta$ a free parameter to be adjusted, with $\beta \in \mathbb{Z_+}$
and $\beta \ge 2$.  Choosing $\beta = 2$ provides a set $S_n^m$ that covers
$\Delta_n^m$ from a certain $n \ge n_0$, where $n_0$ is a value that
increases with $m$. It is possible to bring $n_0$ closer to the origin by
increasing $\beta$, however the extra volume increases as well. What is
interesting is that from $n \ge n_0$, the parallel space is practically
$m!$ times more efficient than a bounding box approach, presenting a great
potential for transforming this space improvement into a performance one.
Studying how parameters $r$ and $\beta$ can be set and relate to each other 
is indeed an interesting open question, since finding the best set becomes 
an optimization problem where the the difference $(1/(r^m) - \beta) - m!$ and 
the term $\beta^{\log_{1/r}(n)}$ are to be minimized.

\section{Discussion}
\label{sec_discussion}
The results from the analysis on recursive GPU mapping for discrete orthogonal
$m$-simplices can serve as a guide for implementing efficient GPU computations
for interaction and simulation problems which are often parallelized using a
bounding-box approach due to its simplicity in implementation.  The $2$-simplex
and $3$-simplex were studied as special cases, re-defining them as a set of
recursive orthotopes. From the analysis it was possible to formulate new $O(1)$
time maps with a potential improvement of $2\times$ and $6\times$ respectively.

The generalization to $m$-simplices presents are greater challenge, as it has
been shown that obtaining an optimal set $S_n^m$ of orthotopes with minimal
extra volume becomes an optimization problem where the scaling and arity
parameters, $r, \beta$ respectively, have to be chosen carefully in order to 
find a small value $n_0$ from which the mapping can take place and obtain a
volume function that introduces a moderate amount of extra volume. Knowing 
what parameters are the optimal for building a recursive set of orthotopes for
any $m$-simplex, as well as provide a rule for the shape of the orthotope
container of $S_n^m$ in any dimension, are indeed interesting questions that 
require further study in order to be answered.

\section*{Acknowledgment}
This project was supported by the research project FONDECYT N$^o$ 3160182 from
CONICYT, as well as by the Nvidia CUDA Research Center at the Department of 
Computer Science (DCC) from University of Chile.



%
\bibliographystyle{plain}
\bibliography{simplex-map}

\begin{thebibliography}{10}

\bibitem{AvrilGA12}
Quentin Avril, Val{\'e}rie Gouranton, and Bruno Arnaldi.
\newblock Fast collision culling in large-scale environments using gpu mapping
  function.
\newblock In {\em EGPGV}, pages 71--80, 2012.

\bibitem{Bedorf:2012:SOG:2133856.2134140}
Jeroen B{\'e}dorf, Evghenii Gaburov, and Simon Portegies~Zwart.
\newblock A sparse octree gravitational n-body code that runs entirely on the
  {GPU} processor.
\newblock {\em J. Comput. Phys.}, 231(7):2825--2839, April 2012.

\bibitem{1054599}
J.~Costello.
\newblock On the number of points in regular discrete simplex (corresp.).
\newblock {\em IEEE Transactions on Information Theory}, 17(2):211--212, Mar
  1971.

\bibitem{ConwaysLife}
M.~Gardner.
\newblock {The fantastic combinations of John Conway's new solitaire game
  ``life''}.
\newblock {\em Scientific American}, 223:120--123, October 1970.

\bibitem{springerlink_gustavson}
Fred Gustavson.
\newblock New generalized data structures for matrices lead to a variety of
  high performance algorithms.
\newblock In Roman Wyrzykowski, Jack Dongarra, Marcin Paprzycki, and Jerzy
  Wasniewski, editors, {\em Parallel Processing and Applied Mathematics},
  volume 2328 of {\em Lecture Notes in Computer Science}, pages 418--436.
  Springer Berlin / Heidelberg, 2006.

\bibitem{1992Huckaby}
Dale~A. Huckaby and Lesser Blum.
\newblock Effect of triplet correlations on the adsorption of a dense fluid
  onto a crystalline surface.
\newblock {\em The Journal of Chemical Physics}, 97(8):5773--5776, 1992.

\bibitem{Ivanov:2007:NPT:1231091.1231100}
Lubomir Ivanov.
\newblock The n-body problem throughout the computer science curriculum.
\newblock {\em J. Comput. Sci. Coll.}, 22(6):43--52, June 2007.

\bibitem{Jung2008}
Jin~Hyuk Jung and Dianne~P. O’Leary.
\newblock Exploiting structure of symmetric or triangular matrices on a gpu.
\newblock Technical report, University of Maryland, 2008.

\bibitem{kepner2011graph}
J.~Kepner and J.~Gilbert.
\newblock {\em Graph Algorithms in the Language of Linear Algebra}.
\newblock Software, Environments, Tools. Society for Industrial and Applied
  Mathematics, 2011.

\bibitem{opencl08}
{Khronos OpenCL Working Group}.
\newblock {\em The OpenCL Specification, version 1.0.29}, 8 December 2008.

\bibitem{Koanantakool:2014:CCP:2683593.2683634}
Penporn Koanantakool and Katherine Yelick.
\newblock A computation- and communication-optimal parallel direct 3-body
  algorithm.
\newblock In {\em Proceedings of the International Conference for High
  Performance Computing, Networking, Storage and Analysis}, SC '14, pages
  363--374, Piscataway, NJ, USA, 2014. IEEE Press.

\bibitem{Li:2010:CME:1955604.1956601}
Qi~Li, Vojislav Kecman, and Raied Salman.
\newblock A chunking method for euclidean distance matrix calculation on large
  dataset using multi-gpu.
\newblock In {\em Proceedings of the 2010 Ninth International Conference on
  Machine Learning and Applications}, ICMLA '10, pages 208--213, Washington,
  DC, USA, 2010. IEEE Computer Society.

\bibitem{5695222}
D.~Man, K.~Uda, H.~Ueyama, Y.~Ito, and K.~Nakano.
\newblock Implementations of parallel computation of euclidean distance map in
  multicore processors and gpus.
\newblock In {\em Networking and Computing (ICNC), 2010 First International
  Conference on}, pages 120--127, 2010.

\bibitem{Man:2011:GIC:2117688.2118809}
Duhu Man, Kenji Uda, Yasuaki Ito, and Koji Nakano.
\newblock A gpu implementation of computing euclidean distance map with
  efficient memory access.
\newblock In {\em Proceedings of the 2011 Second International Conference on
  Networking and Computing}, ICNC '11, pages 68--76, Washington, DC, USA, 2011.
  IEEE Computer Society.

\bibitem{CLEI-2016-navarro}
Crist{\'o}bal~A. Navarro, Benjam{\'in} Bustos, and Nancy Hitschfeld.
\newblock Potential benefits of a block-space {GPU} approach for discrete
  tetrahedral domains.
\newblock In {\em CLEI-2016, XLII Conferencia Latinoamericana de
  Inform{\'a}tica, Valparaiso, Chile, October 10-14, 2016}, 2016.

\bibitem{DBLP:conf/hpcc/NavarroH14}
Cristobal~A. Navarro and Nancy Hitschfeld.
\newblock {GPU} maps for the space of computation in triangular domain
  problems.
\newblock In {\em 2014 {IEEE} International Conference on High Performance
  Computing and Communications, 6th {IEEE} International Symposium on
  Cyberspace Safety and Security, 11th {IEEE} International Conference on
  Embedded Software and Systems, {HPCC/CSS/ICESS} 2014, Paris, France, August
  20-22, 2014}, pages 375--382, 2014.

\bibitem{navhitmat2014}
Cristobal~A. Navarro, Nancy Hitschfeld-Kahler, and Luis Mateu.
\newblock A survey on parallel computing and its applications in data-parallel
  problems using {GPU} architectures.
\newblock {\em Commun. Comput. Phys.}, 15:285--329, 2014.

\bibitem{Nickolls:2010:GCE:1803935.1804055}
John Nickolls and William~J. Dally.
\newblock The gpu computing era.
\newblock {\em IEEE Micro}, 30(2):56--69, March 2010.

\bibitem{nvidia_cuda_guide}
Nvidia-Corporation.
\newblock {\em Nvidia CUDA C Programming Guide}, 2016.

\bibitem{4490127}
J.D. Owens, M.~Houston, D.~Luebke, S.~Green, J.E. Stone, and J.C. Phillips.
\newblock Gpu computing.
\newblock {\em Proceedings of the IEEE}, 96(5):879--899, May 2008.

\bibitem{Ries:2009:TMI:1654059.1654069}
Florian Ries, Tommaso De~Marco, Matteo Zivieri, and Roberto Guerrieri.
\newblock Triangular matrix inversion on graphics processing unit.
\newblock In {\em Proceedings of the Conference on High Performance Computing
  Networking, Storage and Analysis}, SC '09, pages 9:1--9:10, New York, NY,
  USA, 2009. ACM.

\bibitem{Ying:2011:GDD:2065356.2065583}
Zhi Ying, Xinhua Lin, Simon Chong-Wee See, and Minglu Li.
\newblock Gpu-accelerated dna distance matrix computation.
\newblock In {\em Proceedings of the 2011 Sixth Annual ChinaGrid Conference},
  CHINAGRID '11, pages 42--47, Washington, DC, USA, 2011. IEEE Computer
  Society.

\bibitem{DBLP:journals/corr/abs-1108-5815}
Rio Yokota and Lorena~A. Barba.
\newblock Fast n-body simulations on {{GPU}s}.
\newblock {\em CoRR}, abs/1108.5815, 2011.

\end{thebibliography}

\end{document}